\documentclass[twocolumn,showpacs,preprintnumbers,amsmath,amssymb]{revtex4}
\usepackage{dcolumn}
\usepackage{bm}

\usepackage[dvips]{graphicx}
\usepackage{wrapfig,subfigure}
\usepackage{amssymb}
\usepackage{color}

\newcommand{\rz}{\rangle}
\newcommand{\lz}{\langle}

\newcommand{\ha}{\hat{a}}
\newcommand{\had}{\hat{a}^{\dag}}
\newcommand{\im}{\textrm{Im}}

\newcommand{\cn}{\textrm{cn}}
\newcommand{\sn}{\textrm{sn}}
\newcommand{\dn}{\textrm{dn}}
\newcommand{\sd}{\textrm{sd}}
\newcommand{\cd}{\textrm{cd}}
\newcommand{\nd}{\textrm{nd}}
\newcommand{\kp}{\kappa_{+}}
\newcommand{\km}{\kappa_{-}}
\newcommand{\vier}{\!\!\!\!}

\begin{document}

\title{Switching via quantum activation: A parametrically modulated oscillator}

\author{ M. Marthaler and M. I. Dykman}
 \affiliation{
Department of Physics and Astronomy, Michigan State
 University, East Lansing, MI 48824, USA}
\date{\today}

\begin{abstract}
We study switching between period-two states of an underdamped
quantum oscillator modulated at nearly twice its natural
frequency. For all temperatures and parameter values switching
occurs via quantum activation: it is determined by diffusion over
oscillator quasienergy, provided the relaxation rate exceeds the
rate of interstate tunneling. The diffusion has quantum origin and
accompanies relaxation to the stable state. We find the
semiclassical distribution over quasienergy.  For $T=0$, where the
system has detailed balance, this distribution differs from the
distribution for $T\to 0$; the $T=0$ distribution is also
destroyed by small dephasing of the oscillator. The characteristic
quantum activation energy of switching displays a typical
dependence on temperature and scaling behavior near the
bifurcation point where period doubling occurs.
\end{abstract}


\maketitle

\section{Introduction}
\label{sec:Introduction}

Switching between coexisting stable states underlies many phenomena
in physics, from diffusion in solids to protein folding. For
classical systems in thermal equilibrium switching is often
described by the activation law, with the switching probability
being \hbox{$W\propto \exp\left(-\Delta U/kT\right)$}, where $\Delta
U$ is the activation energy. As temperature is decreased, quantum
fluctuations become more and more important, and below a certain
crossover temperature switching occurs via tunneling
\cite{Affleck1981,Grabert1984,Larkin1985}. The behavior of systems
away from thermal equilibrium is far more complicated. Still, for
classical systems switching is often described by an activation type
law, with the temperature replaced by the characteristic intensity
of the noise that leads to fluctuations
\cite{Landauer1962,Ventcel1970,Dykman1979a,Graham1984b,Bray1989,Dykman1990,Maier1993,Kautz1996,Tretiakov2003}.
Quantum nonequilibrium systems can also switch via tunneling between
classically accessible regions of their phase space
\cite{Sazonov1976,Davis1981,Dmitriev1986a,Heller1999}.

In addition to classical activation and quantum tunneling,
nonequilibrium systems have another somewhat counterintuitive
mechanism of transitions between stable states. We call this
mechanism quantum activation and study it in the present paper. It
describes escape from a metastable state due to quantum
fluctuations that accompany relaxation of the system
\cite{Dykman1988a}. These fluctuations lead to diffusion away from
the metastable state and, ultimately, to transitions over the
classical "barrier", that is, the boundary of the basin of
attraction to the metastable state.

\begin{figure}[b]
\begin{center}
\includegraphics[width=3.0in]{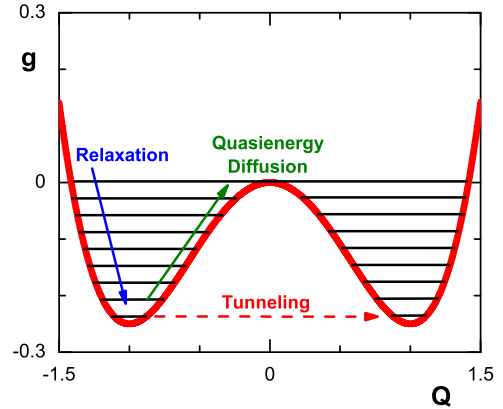}
\caption{(Color online) The effective double-well potential $g(Q)$
of a parametrically modulated oscillator. Sketched are scaled
period-two quasienergy levels (see Sec.~\ref{sec:Dynamics}) in the
neglect of interwell tunneling. The minima of $g$ correspond to
classically stable states of period two motion. The arrows indicate
relaxation, diffusion over quasienergy away from the minima, and
interwell tunneling. The effective Hamiltonian $g(P,Q)$ is defined
by Eq.~(\ref{eq:g}), and $g(Q)\equiv g(P=0,Q)$; the figure refers to
$\mu=0$ in Eq.~(\ref{eq:g}).}\label{fig:D}
\end{center}
\end{figure}

We study quantum activation for periodically modulated systems.
Switching mechanisms for such systems are shown schematically in
Fig.~\ref{fig:D}, where $g(Q)$ is the effective potential of the
system in the rotating frame. This figure describes, in
particular, a nonlinear oscillator studied in the present paper,
which displays period doubling when its frequency is periodically
modulated in time. The energy of periodically modulated systems is
not conserved. Instead they are characterized by quasienergy
$\epsilon$. It gives the change of the wave function
$\psi_{\epsilon}(t)$ when time is incremented by the modulation
period $\tau_F$,
$\psi_{\epsilon}(t+\tau_F)=\exp(-i\epsilon\tau_F/\hbar)\psi_{\epsilon}(t)$,
and is defined modulo $2\pi\hbar/\tau_F$ f\"ur f\'or.

Coupling to a thermal reservoir leads to transitions between the
states of the system. For $T=0$ the transitions are accompanied by
the creation of excitations in the thermal reservoir. The energy
of the system decreases in each transition. However, the
quasienergy may decrease or increase, albeit with different
probabilities $W_{\uparrow}$ and $W_{\downarrow}$. This effect has
quantum origin. It is due to the functions $\psi_{\epsilon}$ being
superpositions of the eigenfunctions $|N\rangle$ of the energy
operator of the system in the absence of modulation. Therefore
bath-induced transitions down in energy $|N\rangle\to |N-1\rangle$
lead to transitions $\psi_{\epsilon} \to \psi_{\epsilon'}$ with
$\epsilon'\lessgtr \epsilon$. The values of $W_{\uparrow}$,
$W_{\downarrow}$ for the latter transitions are determined by the
appropriate overlap integrals and depend on $\epsilon, \epsilon'$.

More probable transitions determine in which direction, with
respect to quasienergy, the system will most likely move. Such
motion corresponds to relaxation over quasienergy.
Fig.~\ref{fig:D} refers to the case $W_{\downarrow}>W_{\uparrow}$.
In this case relaxation corresponds to quasienergy decrease. The
minima of $g(Q)$ are the classical stable states. However, quantum
transitions in which quasienergy increases have a nonzero
probability even for $T=0$. Because this probability is less than
$W_{\downarrow}$, such transitions lead to diffusion over
quasienergy $\epsilon$ away from the minima of $g(Q)$. In turn,
the diffusion leads to a finite-width distribution over $\epsilon$
and ultimately to activated-type overbarrier transitions between
the wells in Fig.~\ref{fig:D}. In fact, discussed in this paper
and sketched in Fig.~\ref{fig:D} are period-two quasienergy
states, with quasienergy $\epsilon$ defined by the condition
$\tilde\psi_{\epsilon}(t+2\tau_F) =
\exp(-2i\epsilon\tau_F/\hbar)\tilde\psi_{\epsilon}(t)$. They are
more convenient for the present problem; their relation to the
standard quasienergy states is explained in
Sec.~\ref{sec:Dynamics}.

Of interest for the problem of switching is the semiclassical
situation where the basins of attraction to the metastable states
(the wells in Fig.~\ref{fig:D}) have a large number $N$ of localized
states. In this case the rate of tunneling decay is exponentially
small. The activation rate is also exponentially small since it is
determined by the ratio of transition probabilities
$W_{\uparrow}/W_{\downarrow}$ raised to the power $N$. Both the
tunneling and activation exponents are $\propto N$ for the situation
sketched in Fig.~\ref{fig:D}. Indeed, the tunneling exponent is
given by the action, in the units of $\hbar$, for classical motion
in the inverted effective potential $-g(Q)$ from one maximum of
$-g(Q)$ to the other. It is easy to see that this action is of order
$N$, unless $g(Q)$ has a special form. Therefore for $N\gg 1$ the
activation exponent is either much larger or much smaller than the
tunneling exponent.

In this paper we develop a theory of the statistical distribution
and consider switching of a parametrically modulated quantum
oscillator. We show that, irrespective of temperature, the
activation exponent is smaller than the tunneling exponent.
Therefore switching always occurs via activation, not tunneling, as
long as the relaxation rate exceeds the tunneling rate.

We study a nonlinear oscillator with frequency modulated at nearly
twice the natural frequency $\omega_0$. When the modulation is
sufficiently strong, the oscillator has two stable vibrational
states with periods $2\tau_F\approx 2\pi/\omega_0$. These states
differ in phase by $\pi$ but otherwise are identical
\cite{LL_Mechanics2004}. They correspond to the minima of the
effective potential in Fig. \ref{fig:D}. The lowest quantized states
in Fig.~\ref{fig:D} are squeezed. Squeezing in a parametric
oscillator has attracted interest in many areas, from quantum optics
\cite{Slusher1985,Wu1986,Loudon1987} to phonons \cite{Garrett1997},
microcantilevers \cite{Rugar1991} and electrons and ions in Penning
traps \cite{Natarajan1994}. Recent progress in systems based on
Josephson junctions and nanoelectromechanical systems
\cite{Siddiqi2004,Claudon2004,Blencowe2004a,Aldridge2005} makes it
possible to study squeezed states in a well controlled and versatile
environment. Both classical and quantum fluctuations can be
investigated and the nature of switching between the states can be
explored. The results can be further used in quantum measurements
for quantum computing, as in the case of switching between
coexisting states of a resonantly driven oscillator
\cite{Siddiqi2004}.

The probability distribution and interstate transitions of a
parametrically modulated oscillator have attracted considerable
attention. Much theoretical work has been done for models where
fluctuations satisfy the detailed balance condition either in the
classical limit \cite{Woo1971} or for $T=0$
\cite{Drummond1981,Wolinsky1988,Drummond1989,Kinsler1991,Kryuchkyan1996}.
Generally this condition does not hold in systems away from
equilibrium. In particular a classical nonlinear parametric
oscillator does not have detailed balance. Switching of such an
oscillator was studied experimentally for electrons in Penning traps
\cite{Tan1993,Lapidus1999}. The measured switching rate
\cite{Lapidus1999} agreed quantitatively with the theory
\cite{Dykman1998}.

\begin{figure}[t]
\begin{center}
\includegraphics[width=3.0in]{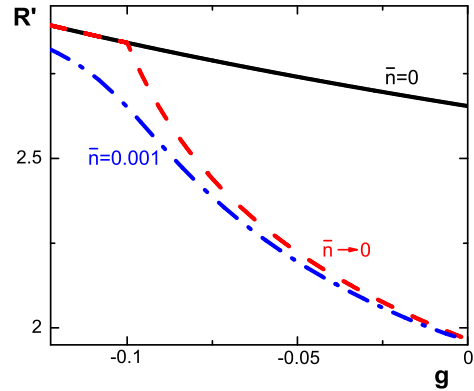}
\caption{(Color online) The scaled inverse temperature $R'\equiv
dR/dg$ of the distribution over scaled period-two quasienergy $g$
of a parametrically modulated oscillator for different oscillator
Planck numbers $\bar n$ in the limit of a large number of
intrawell quasienergy states. The figure refers to $\mu=-0.3$.
}\label{fig:Fr}
\end{center}
\end{figure}

A quantum parametric oscillator also does not have detailed
balance in the general case. The results presented below show that
breaking the special condition where detailed balance holds leads
to a sharp change of the statistical distribution and the
switching rate. This change occurs already for an infinitesimally
small deviation from detailed balance, in the semiclassical limit.
The fragility of the detailed balance solution is previewed in
Fig.~\ref{fig:Fr}. This figure shows the effective inverse
temperature of the intrawell distribution over the scaled
period-two quasienergy $g$. The function $\lambda^{-1}R(g)$ is the
exponent of the distribution $\lambda^{-1}R(g_n)=-\ln\,\rho_n$,
where $\rho_n$ is the population of an $n$-th state and
$\lambda\propto 1/N$ is the scaled Planck constant defined in
Eq.~(\ref{eq:L}) below. The effective inverse temperature
$\lambda^{-1}dR/dg$ depends on $g$ and differs from the inverse
temperature of the bath $T^{-1}$. The $T=\bar{n}=0$ result for
$dR/dg$ is obtained from the solution with detailed balance (here
$\bar n=\left[\exp(\hbar\omega_0/kT)-1\right]^{-1}$ is the Planck
number of the oscillator). It is seen from Fig.~\ref{fig:Fr} that
the $T\rightarrow 0$ ($\bar{n}\to 0$) limit of the solution
without detailed balance does not go over into the $T=0$ result.

\begin{figure}[t]
\begin{center}
\includegraphics[width=3.0in]{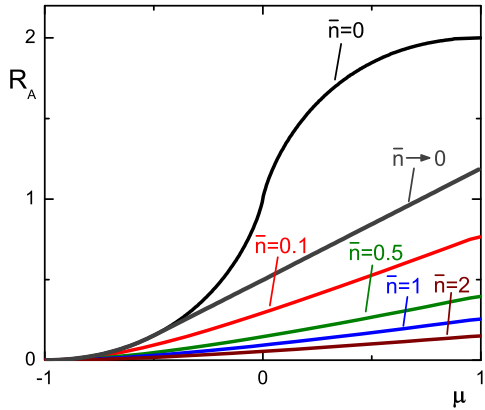}
\caption{(Color online) Quantum activation energy of a transition
between the stable states of period two vibrations of a modulated
oscillator (a phase-flip transition). The transition probability is
$W_{\rm sw} \propto \exp(-R_A/\lambda)$, where $\lambda$,
Eq.~(\ref{eq:L}), is the effective Planck constant. The scaled
switching exponent $R_A$ is plotted as a function of the single
scaled parameter $\mu$ that controls oscillator dynamics,
Eq.~(\ref{eq:mu}), for different values of the oscillator Planck
number $\bar n$. }\label{fig:ActE2}
\end{center}
\end{figure}

The effective quantum activation energy $R_A\approx -\lambda\log
W_{\rm sw}$, which gives the exponent of the switching rate $W_{\rm
sw}$, is shown in Fig.~\ref{fig:ActE2}. It is equal to
$R_A=R(0)-R(g_{\min})$, where $g=0$ and $g=g_{\min}$ are,
respectively, the values of $g(Q)$ at the top and the minimum of the
wells in Fig.~\ref{fig:D}. The $\bar n =0$ detailed balance value for
$R_A$ strongly differs from the $\bar n \to 0$ value in a broad range
of the control parameter $\mu$ that characterizes the detuning of the
modulation frequency from $2\omega_0$, see Eq.~(\ref{eq:mu}) below. It
is also seen from Fig.~\ref{fig:ActE2} that the quantum activation
energy $R_A$ decreases with increasing temperature. Ultimately when
the Planck number of the oscillator becomes large, $\bar{n}\gg 1$, we
have $R_A\propto T^{-1}$, the law of thermal activation, and the
results coincide with the results \cite{Dykman1998} obtained by a
different method.

In Sec.~\ref{sec:Dynamics} we obtain an effective Hamiltonian that
describes the oscillator dynamics in the rotating frame. We
describe the classical dynamics and calculate semiclassical matrix
elements of the coordinate and momentum. In
Sec.~\ref{sec:Balance_equation} we derive the balance equation for
the distribution over quasienergy levels. This distribution is
found in the eikonal approximation in Sec.~\ref{sec:Distribution}.
Its explicit form is obtained in several physically important
limiting cases in Sec.~\ref{sec:Limiting_cases}. In
Sec.~\ref{sec:Switching_exponent} the probability of interstate
switching due to quantum activation is analyzed. In
Sec.~\ref{sec:Fragility} it is shown that the distribution
obtained for $T =0$, where the system has detailed balance, is
fragile: in the limit of a large number of intrawell states it
differs from the distribution for $T\to 0$. It is also
dramatically changed by even weak dephasing due to external noise.
In Sec.~\ref{sec:tunneling} we discuss tunneling between the
coexisting states of period two vibrations and show that
interstate switching occurs via quantum activation rather than
tunneling, if the relaxation rate exceeds the tunneling rate. In
Sec.~\ref{sec:Conclusions} we summarize the model, the
approximations, and the major results.

\section{Dynamics of the parametric oscillator}
\label{sec:Dynamics}

\subsection{The Hamiltonian in the rotating frame}
We will study quantum fluctuations and interstate switching using
an important model of a bistable system, a parametric oscillator.
The Hamiltonian of the oscillator has a simple form
\begin{equation}
\label{eq:H_0(t)}
H_0=\frac{1}{2}p^2+\frac{1}{2}q^2\left(\omega_0^2+F\cos(\omega_F
t)\right)+\frac{1}{4}\gamma q^4\, .
\end{equation}
We will assume that the modulation frequency $\omega_F$ is close to
twice the frequency of small amplitude vibrations $\omega_0$,
and that the driving force $F$ is not large, so that the
oscillator nonlinearity remains small,
\begin{eqnarray}
\label{eq:delta_omega} \delta\omega=\frac{1}{2}\omega_F-\omega_0\,
,\quad
|\delta\omega|\ll\omega_0\, ,\\
\quad F\ll\omega_0^2\, ,\quad |\gamma|\langle q^2 \rangle
\ll\omega_0^2\, .\nonumber
\end{eqnarray}
Here, $\langle q^2 \rangle$ is the mean squared oscillator
displacement; in what follows for concreteness we set $\gamma>0$.

We will change to the rotating frame using the canonical
transformation $U(t)=\exp\left(-i\had\ha\,\omega_Ft/2\right)$,
where $\had$ and $\ha$ are the raising and lowering operators,
\hbox{$\ha=(\hbar\omega_F)^{-1/2}\left(i p+\omega_F q/2\right)$}.
It is convenient to introduce the dimensionless coordinate Q and
momentum P,
\begin{eqnarray}
\label{eq:UdU}
 U^{\dag}(t)q U(t)=C\left[P\cos(\omega_F t/2)-Q\sin(\omega_F t/2)\right]\, ,\\
 U^{\dag}(t)p U(t)=-C\frac{\omega_F}{2}\left[P\sin(\omega_F t/2)+Q\cos(\omega_F
 t/2)\right] ,\nonumber
\end{eqnarray}
where $C=(2F/3\gamma)^{1/2}$. The commutation relation between $P$
and $Q$ has the form
\begin{equation}\label{eq:L}
[P,Q]=-i\lambda\, ,\quad \lambda=3\gamma\hbar/F\omega_F\, .
\end{equation}
The dimensionless parameter $\lambda$ will play the role of
$\hbar$ in the quantum dynamics in the rotating frame. This
dynamics is determined by the Hamiltonian
\begin{equation}
\label{eq:tilde_H_0}
\tilde{H}_0=U^{\dag}H U-i\hbar U^{\dag}\dot{U}\approx
\frac{F^2}{6\gamma}\,\hat{g},
\end{equation}
with
\begin{eqnarray}\label{eq:g}
\hat g & \equiv & g(P,Q)\nonumber \\
& = &
\frac{1}{4}\left(P^2+Q^2\right)^2+\frac{1}{2}(1-\mu)P^2-\frac{1}{2}(1+\mu)Q^2.
\end{eqnarray}
Here we used the rotating wave approximation and disregarded fast
oscillating terms  $\propto \exp(\pm i n\omega_F t)\, ,\,n\geq 1 $.
As a result $\tilde{H}_0$ is independent of time.

\begin{figure}[h]
\begin{center}
\includegraphics[width=3.0in]{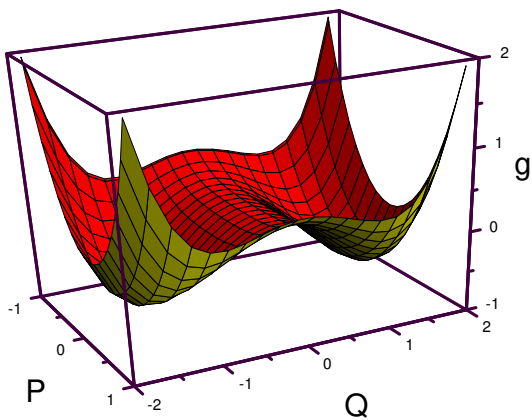}
\caption{(Color online) The scaled effective Hamiltonian of the
oscillator in the rotating frame $g(P,Q)$, Eq.~(\ref{eq:g}), for
$\mu=0.2$. The minima of $g(P,Q)$ correspond to the stable states of
period two vibrations.}\label{fig:g}
\end{center}
\end{figure}

In the "slow" dimensionless time
\begin{equation}
\label{eq:tau}
\tau=t F/2\omega_F
\end{equation}
the Schr\"{o}dinger equation has the form $i\lambda
d\psi/d\tau=\hat{g}\psi$. The effective Hamiltonian $\hat{g}$,
Eq.~(\ref{eq:g}), depends on one parameter
\begin{equation}
\label{eq:mu}
\mu=2\omega_F\delta\omega/F\, .
\end{equation}
For $\mu>-1$, the function $g(P,Q)$ has two minima. They are
located at $P=0$, $Q=\pm(\mu+1)^{1/2}$, and $g_{\rm
min}=-(\mu+1)^2/4$. For $\mu\leq 1$ the minima are separated by a
saddle at $P=Q=0$, as shown in Fig.~\ref{fig:g}. When friction is
taken into account, the minima become stable states of the
parametrically excited vibrations in the classical limit
\cite{LL_Mechanics2004,Dykman1998} [we note that the function $G$
in Ref. \onlinecite{Dykman1998} is similar to $g(P,Q)$, but has
opposite sign].

The function $g(P,Q)$ is symmetric, $g(P,Q)=g(-P,-Q)$. This is a
consequence of the time translation symmetry. The sign change
$(P,Q)\rightarrow(-P,-Q)$ corresponds to the shift in time by the
modulation period $t\rightarrow t+2\pi/\omega_F$, see Eq.
(\ref{eq:UdU}). In contrast to the standard Hamiltonian of a
nonrelativistic particle, the function $g(P,Q)$ does not have the
form of a sum of kinetic and potential energies, and moreover, it is
quartic in momentum P. This structure results from switching to
the rotating frame. It leads to a significant modification of
quantum dynamics, in particular tunneling, compared to the
conventional case of a particle in a static potential.

\subsection{Quasienergy spectrum  and classical
motion}\label{sec:gspec}

\subsubsection{Period-two quasienergy states}

The quasienergies of the parametric oscillator $\epsilon_m$ are
determined by the eigenvalues $g_m$ of the operator $\hat g=
g(P,Q)$, with an accuracy to small corrections $\sim F/\omega_F^2$.
They can be found from the Schr\"odinger equation $\hat g
|m\rz^{(0)}=g_m|m\rz^{(0)}$. Because of the symmetry of $\hat g$ the
exact eigenfunctions $|m\rangle^{(0)}$ are either symmetric or
antisymmetric in $Q$.

The eigenstates $|m\rangle^{(0)}$ give the Floquet states of the
parametric oscillator
$\psi_{\epsilon}(t+\tau_F)=\exp(-i\epsilon\tau_F/\hbar)\psi_{\epsilon}(t)$.
From the form of the unitary operator
$U(t)=\exp(-i\hat{a}^{\dagger}\hat{a}\omega_Ft/2)$ it is clear
that $U(\tau_F)|m\rangle^{(0)}=\pm |m\rangle^{(0)}$, with ``$+$"
for symmetric and ``$-$" for antisymmetric states
$|m\rangle^{(0)}$. Therefore we have $\epsilon_m
=(F^2/6\gamma)g_m$ for symmetric and $\epsilon_m =(F^2/6\gamma)g_m
+ (\hbar\omega_F/2)$ for antisymmetric states.

It is convenient to define period-two quasienergy by the condition
$\tilde\psi_{\epsilon}(t+2\tau_F)=
\exp(-2i\epsilon\tau_F/\hbar)\tilde\psi_{\epsilon}(t)$. In this case
the relation $\epsilon_m =(F^2/6\gamma)g_m$ holds for both symmetric
and antisymmetric states. At the same time, there is simple one to
one correspondence with the standard quasienergy states. Period-two
quasienergy states are particularly convenient for describing
intrawell states $|n\rangle$ for small tunneling between the wells
of $g(P,Q)$. Such states are combinations of symmetric and
antisymmetric states $|m\rangle^{(0)}$. For this reason we use
period-two quasienergies throughout the paper.

\subsubsection{Classical intrawell motion}

We will study the intrawell states $|n\rangle$ in the semiclassical
approximation. The classical equations of motion
\begin{equation}
\label{eq:dQPdt}
\frac{dQ}{d\tau}=\partial_P g\, ,\quad
\frac{dP}{d\tau}=-\partial_Q g\,
\end{equation}
can be explicitly solved in terms of the Jacobi elliptic
functions. The solution is given in Appendix A. There are two
classical trajectories for each $g<0$, one per well of $g(P,Q)$ in
Fig.~\ref{fig:g}. They are inversely symmetrical in phase space
and double periodic in time,
\begin{equation}
\label{eq:periodicity}
 Q(\tau+\tau_p)=Q(\tau)\, ,\quad
P(\tau+\tau_p)=P(\tau)\, ,
\end{equation}
with one real period $\tau_p^{(1)}$ and one complex period
$\tau_p^{(2)} $. This means that
$\tau_p=n_1\tau_p^{(1)}+n_2\tau_p^{(2)}$, where $n_{1,2}=0,\pm
1,\ldots $, and
\begin{eqnarray}
\label{eq:Period}
\tau_p^{(1)}=2^{1/2}|g|^{-1/4}K(m_J)\,
,\\
\tau_p^{(2)}=i\,2^{1/2}|g|^{-1/4}K'(m_J).\nonumber
\end{eqnarray}
Here $K(m_J)$ is the complete elliptic integral of the first kind,
$K'(m_J)=K(1-m_J)$, and $m_J\equiv m_J(g)$,
\begin{equation}
\label{eq:m}
m_J(g)=\frac{\left(\mu+1-2|g|^{1/2}\right)\left(\mu-1+2|g|^{1/2}\right)}{8|g|^{1/2}}-i0
\end{equation}
is the parameter \cite{Abramowitz1972}. The real part of $m_J$ can
be positive or negative; for ${\rm Re}\,m_J<0$, the period
$\tau_p^{(2)}$ has not only imaginary,  but also a nonzero real
part.

The vibration frequency is $\omega(g)=2\pi/\tau_p^{(1)}\equiv
2\pi/\tau_p^{(1)}(g)$. It monotonically decreases with increasing
$g$ in the range where $g(P,Q)$ has two wells, $-(\mu+1)^2/4<g<0$,
and $\omega(g)\rightarrow 0$ for $g\rightarrow 0$, i.e. for $g$
approaching the saddle-point value. The periodicity of $Q(\tau)$,
$P(\tau)$ in imaginary time turns out to be instrumental in
calculating matrix elements of the operators $P$, $Q$ on the
intrawell wave functions $|n\rangle$ as discussed in the next
section.

\subsection{The semiclassical matrix elements}
Of central interest to us will be the case where the number of
states with $g_n<0$, that is the number of states inside the wells
of $g(P,Q)$ in Fig.~\ref{fig:g}, is large. Formally, this case
corresponds to the limit of a small effective Planck constant
$\lambda$. For $\lambda\ll 1$ the wave function $|n\rz$ in the
classically accessible region can be written in the form
\begin{eqnarray}
\label{eq:Psi} |n\rz=C'(\partial_P
g)^{-1/2}\exp\left[i\,S_n(Q)/\lambda\right]+{\rm c.c.}\,
,\\
S_n(Q)=\int_{Q_{n}}^{Q}P(Q',g_n)dQ'\, .\nonumber
\end{eqnarray}
Here, $C'$ is the normalization constant, $P(Q,g_n)$ is the momentum
for a given $Q$ as determined by the equation $g(P,Q)=g_n$, $Q_{n}$
is the classical turning point, $P(Q_{n},g_n)=0$, and the derivative
$\partial_P g$ is calculated for $P=P(Q,g_n)$.

We parameterize the classical trajectories (\ref{eq:dQPdt}) by their
phase $\phi=\omega(g)\tau$, which we count (modulo $2\pi$) from its
value at $Q_{n}$. With this parametrization we have
\begin{equation}
\label{eq:dS} S_{n+m}(Q)-S_n(Q)\approx \lambda m\phi\, ,
\end{equation}
for $|m|\ll\lambda^{-1}$; on a classical trajectory, $Q$ is an even
and $P$ is an odd function of $\phi$.

With Eqs.~(\ref{eq:Psi}) and (\ref{eq:dS}), we can write the
semiclassical matrix element of the lowering operator
$\ha=(2\lambda)^{-1/2}(P-iQ)$ as a Fourier component of
$a(\phi;g)= (2\lambda)^{-1/2}[P(\tau;g)-iQ(\tau;g)]$ over the
phase $\phi=\omega(g)\tau$ \cite{LL_QM81},
\begin{eqnarray}
\label{eq:a}
\lz n+m|\hat a|n\rz & \equiv & a_m(g_n), \\
a_m(g) & = & \frac{1}{2\pi}\int_0^{2\pi}d\phi \exp(-i m\phi) a
(\phi;g)\, , \nonumber
\end{eqnarray}
to lowest order in $\lambda$. As we discuss below, these matrix
elements determine relaxation of the oscillator.

We will be interested in calculating $a_m(g)$ for $g<0$. Then if
we neglect interwell tunneling, we have two sets of wave functions
$|n\rz$, one for each well. The interwell matrix elements of the
operator $\hat a$ are exponentially small. The intrawell matrix
elements are the same for the both wells except for the overall
sign, because $Q$ and $P$ in different wells have opposite signs
on the trajectories with the same $g$. We will consider the matrix
elements $a_m(g_n)$ for the states in the right well $Q>0$.

\begin{figure}[t]
\begin{center}
\includegraphics[width=3.0in]{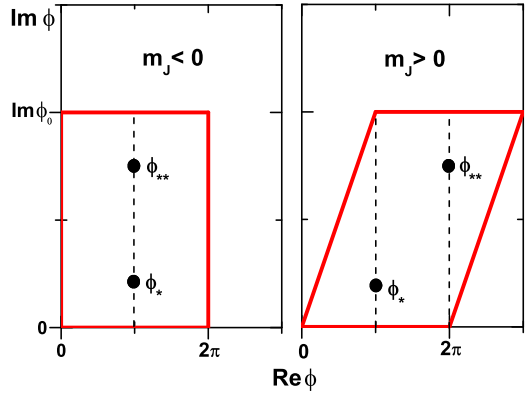}
\caption{(Color online) The contours of integration in the complex
phase plane for calculating Fourier components of the oscillator
coordinate $Q$ and momentum $P$. The left and right panels refer,
respectively, to the negative and positive real part of the
parameter of the elliptic functions $m_J$, Eq.~(\ref{eq:m}) [the
imaginary part of $m_J$ is infinitesimally small]. Both $Q$ and $P$
have poles at $\phi=\phi_*,\phi_{**}$, but $a(\phi;g)$ has a pole
only for $\phi=\phi_*$. The increment of phase by $\phi_0$ leads to
a transition between the trajectories with a given $g$ in different
wells of $g(P,Q)$.} \label{fig:CL}
\end{center}
\end{figure}

The integral (\ref{eq:a}) can be evaluated using the double
periodicity of the functions $Q$, $P$. It is clear that
$a(\phi+2\pi, g)=a(\phi;g)$ for any complex $\phi$. In addition,
from Eq.~(\ref{eq:Pe}) for $g<0$ we have $a(\phi+\phi_0,
g)=-a(\phi;g)$, where $\phi_0=\pi(1+\tau_p^{(2)}/\tau_p^{(1)})$ is a
complex number, which is determined by the ratio of the periods of
$Q$ and $P$. The shift in time by $\phi_0/\omega(g)$ corresponds to
a transition from $Q(\tau),P(\tau)$ in one well to $Q(\tau),P(\tau)$
in the other well. We can now write the matrix element $a_m(g)$ as
\begin{equation}
\label{eq:Cint}
a_m(g)=\frac{1}{2\pi}\left[1+e^{-im\phi_0}\right]^{-1}\oint_C
d\phi \,e^{-im\phi}a(\phi;g)\, ,
\end{equation}
where integration is done along the contour $C$ in Fig.
\ref{fig:CL}. It is explained in Appendix \ref{ap:CM} that each of
the functions $Q\left(\phi/\omega(g)\right)$,
$P\left(\phi/\omega(g)\right)$ has two poles inside the contour
$C$ \cite{Abramowitz1972}. Therefore it is easy to calculate the
contour integral (\ref{eq:Cint}). The result has a simple form
\begin{equation}
\label{eq:am}
a_m(g)=-i(2\lambda)^{-1/2}\omega(g)\frac{\exp(-im\phi_*)}{1+\exp(-im\phi_0)}\,
,
\end{equation}
with $\phi_*$ given by the equation
\begin{eqnarray}
\label{eq:TS}
\cn(2K\phi_*/\pi)=-\left(\frac{1+\mu+2|g|^{1/2}}{1+\mu-2|g|^{1/2}}\right)^{1/2},\\
\im\,\phi_*<\frac{\pi}{2}\im\,\left[\tau_p^{(2)}/\tau_p^{(1)}\right]
=\frac{1}{2}\im\,\phi_0. \nonumber
\end{eqnarray}

An important feature of the matrix elements $a_m(g)$ is their
exponential decay for large $|m|$. From Eqs.~(\ref{eq:am}) and
(\ref{eq:TS})
\begin{eqnarray}
\label{eq:aE}
a_m(g)\propto\exp\left[-m\,\im(\phi_0-\phi_{*})\right]\, ,\quad m\gg 1\, ,\\
a_m(g)\propto\exp[-|m|\im\,\phi_*]\, ,\quad -m\gg 1\, .\nonumber
\end{eqnarray}
We note that the decay is asymetric with respect to the sign of $m$.
This leads to important features of the probability distribution of
the oscillator.

\section{Balance equation}
\label{sec:Balance_equation}

Coupling of the oscillator to a thermal reservoir leads to its
relaxation. We will first consider the simplest type of relaxation.
It arises from coupling linear in the oscillator coordinate $q$ and
corresponds to decay processes in which the oscillator makes
transitions between neighboring energy levels, with energy $\approx
\hbar\omega_0$ transferred to or absorbed from the reservoir. We
will assume that the oscillator nonlinearity is not strong and that
the detuning $\delta\omega$ of the modulation frequency is small,
whereas the density of states of the reservoir weighted with
interaction is smooth near $\omega_0$. Then the quantum kinetic
equation for the oscillator density matrix $\rho$ in the rotating
frame has the form
\begin{eqnarray}
\label{eq:Master}
\frac{\partial\rho}{\partial\tau}=i\lambda^{-1}[\rho,g]-\hat{\eta}\rho\, ,\\
\hat{\eta}\rho=\eta\left[(\bar{n}+1)(\had\ha\rho-2\ha\rho\had+
\rho\had\ha)\right.\nonumber\\
\left.+\bar{n}(\ha\had\rho-2\had\rho\ha +\rho\ha\had)\right]\,
,\nonumber
\end{eqnarray}
where $\eta$ is the dimensionless relaxation constant and
$\bar{n}=[\exp\left(\hbar\omega_0/k T\right)-1]^{-1}$ is the Planck
number.

We will assume that relaxation is slow so that the broadening of
quasienergy levels is much smaller than the distance between them,
$\eta\ll \omega(g)$. Then off-diagonal matrix elements of $\rho$ on
the states $|n\rz$ are small. We note that, at the same time,
off-diagonal matrix elements of $\rho$ on the Fock states of the
oscillator $|N\rz$ do not have to be small.

To the lowest order in $\eta/\omega(g)$ relaxation of the diagonal
matrix elements $\rho_n=\lz n|\rho|n\rz$, is described by the
balance equation
\begin{equation}
\label{eq:BE} \frac{\partial\rho_n}{\partial
\tau}=-2\eta\sum_{n'}\left(W_{nn'}\rho_n-W_{n'n}\rho_{n'}\right)\, ,
\end{equation}
with dimensionless transition probabilities
\begin{equation}
\label{eq:WMN}
W_{nn'}=(\bar{n}+1)|\lz n'|\ha|n \rz|^2+\bar{n}|\lz
n|\ha|n'\rz|^2\, .
\end{equation}

It follows from Eqs.~(\ref{eq:am}) and (\ref{eq:WMN}) that, even
for $T=0$, the oscillator can make transitions to states with both
higher and lower $g$, with probabilities $W_{nn'}$ with $n'>n$ and
$n'<n$, respectively. This is in spite the fact that transitions
between the Fock states of the oscillator for $T=0$ are only of
the type $|N\rz\rightarrow|N-1\rz$.

The explicit expression for the matrix elements (\ref{eq:am}) makes
it possible to show that the probability of a transition to a lower
level of $\hat g$ is larger than the probability of a transition to
a higher level, that is, $W_{n'n}>W_{nn'}$ for $n'>n$. Therefore the
oscillator is more likely to move down to the bottom of the
initially occupied well in Fig.~\ref{fig:g}. This agrees with the
classical limit in which the stable states of an underdamped
parametric oscillator are at the minima of $g(P,Q)$. However, along
with the drift down in the scaled period-two quasienergy $g$, even
for $T=0$ there is also diffusion over quasienergy away from the
minima of $g(P,Q)$, due to nonzero transition probabilities
$W_{n\,n'}$ with $n'>n$.

Equation (\ref{eq:BE}) must be slightly modified in the presence
of tunneling between the states with $g_n<0$. The modification is
standard. One has to take into account that the matrix elements of
$\rho$ depend not only on the number $n$ of the period-two
quasienergy level inside the well, but also on the index $\alpha$
which takes on two values $\alpha=\pm 1$, that specify the wells
of $g(P,Q)$. These values can be associated with the eigenvalues
of the pseudospin operator $\sigma_{z}$. The matrix elements of
the operator $\hat{\eta}\rho$ on the wave functions of different
wells are exponentially small and can be disregarded. The operator
describing interwell tunneling can be written in the pseudospin
representation as
$-i(2\lambda)^{-1}T(g)\left[\sigma_x,\rho\right]$, where $T(g)$ is
the tunneling splitting of the states in different wells.

We assume that the tunneling splitting is much smaller than $\eta$.
Then after a transient time $\tau\sim\eta^{-1}$, there is formed a
quasistationary distribution over the states $|n\rz$ inside each of
the wells of $g(P,Q)$ in Fig.~\ref{fig:g}. This distribution can be
found from Eq.~(\ref{eq:BE}) using the wave function $|n\rz$
calculated in the neglect of tunneling. Interwell transitions occur
over much longer time.

\section{Distribution over intrawell states}
\label{sec:Distribution}

The stationary intrawell probability distribution can be easily
found from Eqs.~(\ref{eq:BE}) and (\ref{eq:WMN}) if the number of
levels with $g_n<0$ is small. Much more interesting is the
situation where this number is large. It corresponds to the limit
of small $\lambda$. We will be interested primarily in the
quasistationary distribution over the states in the well in which
the system was initially prepared. It is formed over time $\sim
\eta^{-1}$ and is determined by setting the right hand side of
Eq.~(\ref{eq:BE}) equal to zero. The resulting equation describes
also the stationary distribution in both wells, which is formed
over a much longer time given by the reciprocal rate of interwell
transitions.

For $\lambda\ll 1 $ we can use the Wentzel-Kramers-Brillouin (WKB)
approximation both to calculate the matrix elements in $W_{nn'}$
(\ref{eq:WMN}) and to solve the balance equation. The solution
should be sought in the eikonal form
\begin{equation}
\label{eq:Rho} \rho_n=\exp\left[-R_n/\lambda\right]\, ,\quad
R_n=R(g_n).
\end{equation}
It follows from Eq.~(\ref{eq:aE}) that the transition probabilities
$W_{n+m\,n}$ rapidly decay for large $|m|$. Therefore in the balance
equation (\ref{eq:BE}) we can set
\begin{eqnarray}\label{eq:RP}
&&\rho_{n+m}\approx\rho_n\exp\left[-m\omega(g_n)R'(g_n)\right]\, ,\\
&& R'(g)=dR/dg.\nonumber
\end{eqnarray}
The corrections to the exponent in Eq.~(\ref{eq:RP}) of order
$\lambda m^{2}\omega^{2}R''$, $\lambda m^{2} \omega\omega' R' $ are
small for $\lambda\ll 1$ (prime indicates differentiation over g).
We note that $\omega(g)R'(g)$ is not small in the quantum regime and
the exponential in (\ref{eq:RP}) will not be expanded in a series in
$R'$.

From Eq.~(\ref{eq:BE}), the function $R'(g)$ is determined by the
polynomial equation
\begin{eqnarray}
\label{eq:BEP} &&\sum_m W_{n+m\,n}\left(1-\xi^{m}\right)=0,\\
&&\xi=\exp\left[-\omega(g_n)R'(g_n)\right].\nonumber
\end{eqnarray}
Here the sum goes over positive and negative $m$. In obtaining
Eq.~(\ref{eq:BEP}) we have used the relation $W_{n+m\, n}=W_{n\,
n-m}$ for $|m|\ll n$, which is the consequence of the WKB
approximation for the matrix elements: $\lambda\omega(g)
d\ln\left[a_m(g)\right]/dg\ll 1$ for $\lambda\ll 1$.

Equation (\ref{eq:BEP}) has a trivial solution $\xi=1$, which is
unphysical. Because all coefficients $W_{n+m\,n}$ are positive and
$W_{n+m\, n}>W_{n-m\,n}$ for $m>0$, one can show that the
polynomial in the left hand side of Eq.~(\ref{eq:BEP}) has one
extremum in the interval $0<\xi<1$. Since its derivative for
$\xi=1$ is negative and it goes to $-\infty$ for $\xi\rightarrow
0$, it has one root in this interval. This root gives the value
$R'(g_n)$. Since $\xi < 1$, we have $R'(g)>0$. In turn $R'(g)$
gives $R(g)$ and thus the distribution $\rho_n$. In obtaining
$R(g)$ from $R'(g)$, in the spirit of the eikonal approximation,
one should set $R(g_{\min}) =0$, where $g_{\min}=-(\mu+1)^2/4$ is
the minimal quasienergy.

\begin{figure}[b]
\begin{center}
\includegraphics[width=3.0in]{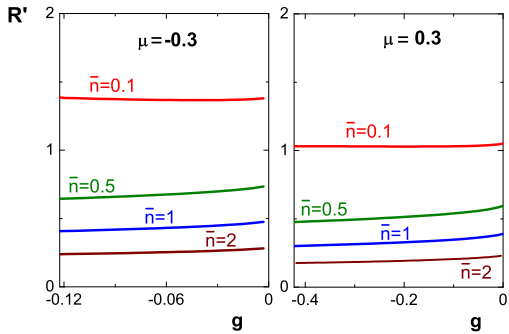}
\caption{(Color online) The scaled inverse temperature
$R^{\prime}$ of the distribution over scaled period-two
quasienergy $g$ for two values of the control parameter $\mu$ and
for different oscillator Planck numbers $\bar n$. For $\bar
n\gtrsim 0.1$ the function $R^{\prime}$ only weakly depends on $g$
inside the well, $g_{\min}< g < 0$.}\label{fig:dSdg}
\end{center}
\end{figure}

It is seen from Eq.~(\ref{eq:Rho}) that $\lambda^{-1}R'(g)$ has
the meaning of the effective inverse temperature of the
distribution over period-two quasienergy. The function $R'(g)$ can
be calculated for different values of the control parameter $\mu$
and different oscillator Planck numbers $\bar{n}$ by solving Eq.
(\ref{eq:BEP}) numerically. The numerical calculation is
simplified by the exponential decay of the coefficients $W_{n+m\,
n}$ with $|m|$. The results are shown in Fig.~\ref{fig:dSdg}. The
function $R'(g)/\lambda$ smoothly varies with $g$ in the whole
range $g_{\min}\leq g<0$ of intrawell values of $g$, except for
very small $\bar{n}$.

\begin{figure}[h]
\includegraphics[width=3.0in]{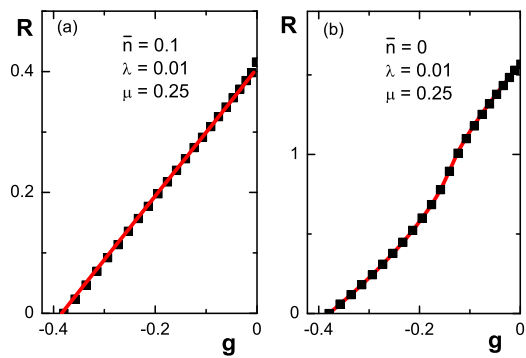}
\caption{(Color online) Comparison of the results of the eikonal
approximation for the scaled logarithm of the probability
distribution $R$ over period-two quasienergy $g$ (solid lines) with
the results obtained by direct calculation of the transition
probabilities followed by numerical solution of the balance equation
(squares).} \label{fig:Num}
\end{figure}

Numerical results on the logarithm of the distribution
$R(g_n)=-\lambda \ln \rho_n$ obtained from Eq.~(\ref{eq:BEP}) for
$\bar n=0.1$ are compared in Fig~\ref{fig:Num}(a) with the results
of the full numerical solution of the balance equation
(\ref{eq:BE}). In this latter calculation we did not use the WKB
approximation to find the transition probabilities $W_{n\,n'}$.
Instead they were obtained by solving numerically the
Schr\"{o}dinger equation $\hat g|n\rz=g_n|n\rz$ and by calculating
$W_{n\,n'}$ as the appropriately weighted matrix elements
$\bigl|\langle n|\hat a|n'\rangle\bigr|^2 $, Eq.~(\ref{eq:WMN}).

In the calculation we took into account that, for small $\lambda$,
the levels of $\hat g$ form tunnel-split doublets. The tunneling
splitting is small compared to the distance between intrawell states
$\lambda\omega(g)$. The exact eigenfunctions of the operator $\hat
g$ are well approximated by symmetric and antisymmetric combinations
of the intrawell wave functions. This allows one to restore the
intrawell wave functions from the full numerical solution and to
calculate the matrix elements $W_{n\,n'}$. By construction, such
numerical approach gives $R(g)$ only for the values of $g$ that
correspond to quasienergy levels $g_n$. It is seen from
Fig.~\ref{fig:Num}(a) that the two methods give extremely close
results for small $\lambda$ (see also below).

\section{The distribution in limiting cases}
\label{sec:Limiting_cases}

The effective inverse temperature $R'(g)/\lambda$ and the
distribution $\rho_n$ can be found in several limiting cases. We
start with the vicinity of the bottom of the wells of $g(P,Q)$. Here
classical vibrations of $P$, $Q$ are nearly harmonic. To leading
order in $\delta g=g-g_{\min}$ we have $\lz n+m|
\hat{a}|n\rz\propto\delta_{|m|,1}\delta g^{1/2}$ for $m\neq 0$, and
therefore the transition probabilities $W_{n+m\, n}\propto
\delta_{|m|,1}$. Then Eq.~(\ref{eq:BEP}) becomes a quadratic
equation for $\xi$, giving
\begin{eqnarray}
\label{eq:RpA1}
R'(g_{\min}) & = &
\frac{1}{2}(\mu+1)^{-1/2}\\
& & \times\ln\frac{(\mu+2)(2\bar{n}+1)+2\sqrt{\mu+1}}
{(\mu+2)(2\bar{n}+1)-2\sqrt{\mu+1}}. \nonumber
\end{eqnarray}
The inverse effective temperature $R'(g_{\min})$ as given by
Eq.~(\ref{eq:RpA1}) monotonically decreases with the increasing
Planck number $\bar{n}$, i.e., with increasing bath temperature
$T$. We note that $R'(g_{\min})$ smoothly varies with $\bar{n}$
for low temperatures $\bar{n}\ll 1$, except for small $|\mu|$. We
have $\partial R'(g_{\min})/\partial \bar{n}=-2^{2/3}(\mu+1)/\mu$
for $\bar n\to 0$. For $\mu=0$, on the other hand, we have
$R'(g_{\min})\approx \hbar\omega_F/4 k T $, i.e. the effective
inverse temperature for $g=g_{\min}$ is simply $\propto T^{-1}$.

We now consider the parameter ranges where $R'(g)$ can be found for
all $g$.

\subsection{Classical limit}

For $\bar{n}\gg 1$ the transition probabilities become nearly
symmetric, $\left|W_{n+m\,n}-W_{n-m\,n}\right|\ll W_{n+m\,n}$. As a
result, the effective inverse temperature $R'/\lambda$ becomes
small, and the ratio $\rho_{n+m}/\rho_n$ can be expanded in
$R'(g_n)/\lambda$. This gives
\begin{equation}
\label{eq:CW} R'(g)=2\omega^{-1}(g)\sum_m m W_{n+m\,n}/\sum_m
m^2W_{n+m\, n}\,.
\end{equation}
It is shown in Appendix \ref{ap:C} that Eq.~(\ref{eq:CW}) can be
written in the simple form
\begin{eqnarray}
\label{eq:RC}
R'(g)=\frac{2}{2\bar{n}+1}M(g)/N(g)\, ,\\
M(g)=\int\vier\int_{A(g)}dQ dP\, ,\nonumber\\
N(g)=\frac{1}{2}\int\vier\int_{A(g)} dQ dP(\partial_Q^2
g+\partial_P^2 g)\, ,\nonumber
\end{eqnarray}
where the integration is performed over the area $A(g)$ of the phase
plane (Q,P) encircled by the classical trajectory $Q(\tau)$,
$P(\tau)$ with given $g$.

From Eq.~(\ref{eq:RC}), the effective inverse temperature
$R'/\lambda$ is $\propto \left[\lambda(2\bar{n}+1)\right]^{-1}$.
In the high-temperature limit $(2\bar{n}+1)\approx
2kT/\hbar\omega_0$. Therefore $R'/\lambda$ is $\propto T^{-1}$ and
does not contain $\hbar$, as expected. Eq.~(\ref{eq:RC}) in this
limit coincides with the expression for the distribution obtained
in Ref.~\onlinecite{Dykman1998} using a completely different
method.

\subsection{Vicinity of the bifurcation point}

The function $R'(g)$ may be expected to have a simple form for
$\mu$ close to the bifurcation value $\mu_B=-1$ where the two
stable states of the oscillator merge together and $g(P,Q)$
becomes single well. This is a consequence of the universality
that characterizes the dynamics near bifurcation points and is
related to the slowing down of motion and the onset of a soft
mode. The situation we are considering here is not the standard
situation of the classical theory where the problem is reduced to
fluctuations of the soft mode. The oscillator is not too close to
the bifurcation point, its motion is not overdamped and the
interlevel distance exceeds the level broadening. Still as we show
$R(g)$ has a simple form.

The motion slowing down for $\mu$ approaching $-1$ leads to the
decrease of the vibration frequencies. From Eq.~(\ref{eq:Period}),
\[\omega(g)=2\pi/\tau_p^{(1)}\propto|g|^{1/4}\leq
\left[\left(\mu+1\right)/2\right]^{1/2}\]
for $\mu+1\ll 1$. Therefore one may expect that $\omega(g) R'(g)$
becomes small, and we can again expand $\rho_{n+m}/\rho_n$ in
$\omega(g_n) R'(g_n)$, as in Eq.~(\ref{eq:CW}). One can justify
this expansion more formally by noticing that the transition
probabilities $W_{n+m\, n}$ are nearly symmetric near the
bifurcation point $\left|(W_{n+m\, n}-W_{n-m\, n})\right|\ll
W_{n-m\, n}$. In the present case the latter inequality is a
consequence of the relation $\im\left(\phi_0-2\phi_*\right)\ll 1$
in Eq.~(\ref{eq:am}), which leads to
$\bigl|\left|a_{-m}(g)\right|-\left|a_{m}(g)\right|\bigl|\ll
|a_{m}(g)|$. In turn, the above relation between $\phi_0$ and
$\phi_*$ can be obtained from Eqs.~(\ref{eq:m}) and (\ref{eq:TS}),
which show that the right-hand side of Eq.~(\ref{eq:TS}) is
$\approx -\left|m_J/\left(1-m_J\right)\right|^{1/2}$ for $\mu+1\ll
1$. Since
$\left|\cn\left(K+i\,K'\right)\right|=\left|m_J/\left(1-m_J\right)\right|^{1/2}$
\cite{Abramowitz1972}, we have from Eqs.~(\ref{eq:Period}) and
(\ref{eq:TS}) $\im\,\phi_*\approx \im\,
\tau_p^{(2)}/2\tau_p^{(1)}=\im\, \phi_0/2$.

It follows from the above arguments that near the bifurcation point
$R'(g)$ is given by Eq.~(\ref{eq:RC}) for arbitrary Planck number
$\bar{n}$, i.e., for arbitrary temperature. Since $\partial_Q^2
g+\partial _P^2 g\approx 2$ for $\mu+1\ll 1$, we obtain from
Eq.~(\ref{eq:RC}) a simple explicit expression
\begin{equation}\label{eq:Rc0}
R'(g)\approx 2/(2\bar{n}+1)\, ,
\end{equation}
Eq.~(\ref{eq:Rc0}) agrees with Eq.~(\ref{eq:RpA1}) near $g_{\min}$
in the limit $\mu+1\ll 1$. It shows that the inverse effective
temperature $R'/\lambda$ is independent of $g$. It monotonically
decreases with increasing temperature $T$.

\subsection{Zero temperature: detailed balance}

The function $R'(g)$ can be also obtained for $T=\bar{n}=0$. This
is a consequence of detailed balance that emerges in this case
\cite{Kryuchkyan1996}. The detailed balance condition is usually a
consequence of time reversibility, which does not characterize the
dynamics of a periodically modulated oscillator. So, in the
present case detailed balance comes from a special relation
between the parameters for $T=0$. Detailed balance means that
transitions back and forth between any two states are balanced. It
is met if the ratio of the probabilities of direct transitions
between two states is equal to the ratio of probabilities of
transitions via an intermediate state
\begin{eqnarray}
\label{eq:detailed_balance_condition} \frac{W_{n\,n'}}{W_{n'\,n}}=
\frac{W_{n\,n''}\,W_{n''\,n'}}{W_{n'\,n''}\,W_{n''\,n}}.
\end{eqnarray}

One can see from Eqs.~(\ref{eq:am}) and (\ref{eq:WMN}) that the
condition (\ref{eq:detailed_balance_condition}) is indeed met for
$\bar{n}=0$. Therefore the balance equation has a solution
$\rho_{n}/\rho_{n'}=W_{n'n}/W_{n\,n'}$, which immediately gives
\begin{equation}
\label{eq:R0} R'(g)=2\omega^{-1}(g)\im\left(\phi_0-2\phi_*\right).
\end{equation}

\begin{figure}[h]
\begin{center}
\includegraphics[width=3.0in]{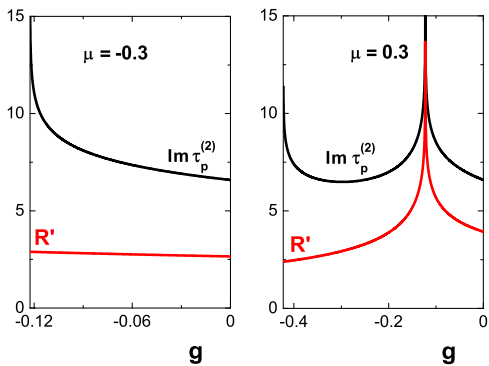}
\caption{(Color online) The scaled inverse temperature $R'$ and the
imaginary part of the period $\tau_p^{(2)}$ of the oscillator
coordinate and momentum $Q(\tau;g), P(\tau;g)$ for $T=0$. For $\mu >
0$ both $R'$ and ${\rm Im}\,\tau_p^{(2)}$ have a logarithmic
singularity.}\label{fig:Randt}
\end{center}
\end{figure}

The function $R'(g)$ is plotted in Fig.~\ref{fig:Randt} for two
values of the control parameter $\mu$. It displays different
behavior depending on the sign of $\mu$. For $\mu<0$ the parameter
of the elliptic function $m_J$, Eq.~(\ref{eq:m}),  is negative for
all $g<0$. Therefore the periods $\tau_p^{(1,2)}(g)$ of $P(\tau)$,
$Q(\tau)$, Eq.~(\ref{eq:Period}), are smooth functions of $g$,
except near $g_{\min}=-\frac{1}{4}(\mu+1)^2$, where $\im\,
\tau_p^{(2)}\approx (1+\mu)^{-1/2}\left|\ln \left(g-g_{\min}\right)
\right|$. This divergence of  $\im\, \tau_p^{(2)}$ is seen in
Fig.~\ref{fig:Randt}. The function $R'(g)$ remains finite near
$g_{\min}$, Eq.~(\ref{eq:RpA1}).

For $\mu>0$, on the other hand, the function $R'(g)$ has a
singularity. Its location $g_d$ is determined by the condition
$m_J=0$, which gives $g_d=-(1-\mu)^2/4$. For small $|g-g_d|$ we
have $\im\, \phi_0\propto
\im\,\tau_p^{(2)}\propto\left|\ln\left(|g-g_d|\right)\right|$,
whereas $\phi_*$, Eq.~(\ref{eq:TS}) remains finite for $g=g_d$.
Therefore $R'(g)$ diverges logarithmically at $g_d$, as seen from
Fig.~\ref{fig:Randt}(b).

Physically the divergence of the inverse temperature is related to
the structure of the transition probabilities $W_{n\,n'}$ for
$\bar{n}=0$. For $g\rightarrow g_d$ we have $W_{n\,n+m}\propto
(g_n-g_d)^{2m}$ for $m>0$, see Eqs.~(\ref{eq:am}) and
(\ref{eq:WMN}). This means that, for $g_n$ close to $g_d$,
diffusion towards larger $g$ slows down. The slowing down leads to
a logarithmic singularity of $R'(g)$. The eikonal approximation is
inapplicable for $g$ close to $g_d$. However, the width of the
range of $g$ where this happens is small, $\delta g \sim \lambda$,
as follows from the discussion below Eq.~(\ref{eq:RP}). In
addition, there are corrections to the balance equation due to
off-diagonal terms in the full kinetic equation. These corrections
give extra terms $\propto \eta^2/\omega^2(g_d)$ in the transition
probabilities $W_{n\,n'}$. The analysis of these corrections as
well as features of $R$ that are not described by the eikonal
approximation is beyond the scope of this paper because, as we
show, these features are fragile.

The function $R(g_n)=-\lambda\ln \rho_n$ obtained by integrating
Eq.~(\ref{eq:R0}) is compared in Fig.~\ref{fig:Num}(b) with the
result of the numerical solution of the balance equation
(\ref{eq:BE}) with numerically calculated transition probabilities
$W_{n\,n'}$. The semiclassical and numerical results are in
excellent agreement. We checked that the agreement persists for
different values of the control parameter $\mu$ and for almost all
$\lambda \ll 1$, except for a few extremely narrow resonant bands of
$\lambda$.

\section{Switching exponent}
\label{sec:Switching_exponent}

Quantum diffusion over quasienergy described by Eq.~(\ref{eq:BE})
leads to switching between the classically stable states of the
oscillator at the minima of $g(P,Q)$ in Fig.~\ref{fig:g}. The
switching rate $W_{\rm sw}$ is determined by the probability to
reach the top of the barrier of $g(P,Q)$, that is by the
distribution $\rho_n$ for such $n$ that $g_n=0$. To logarithmic
accuracy
\begin{equation}
\label{eq:W_sw} W_{\rm sw}=C_{\rm sw}\times
\exp\left(-R_A/\lambda\right),\quad R_A=\int_{g_{\min}}^{0}R'(g)dg,
\end{equation}
where $R'(g)$ is given by Eq.~(\ref{eq:BEP}). The parameter $C_{\rm
sw}$ is of the order of the relaxation rate $\eta$ due to coupling
to a thermal bath.

The quantity $R_A$ plays the role of the activation energy of
escape. The activation is due to quantum fluctuations that accompany
relaxation of the oscillator, and we call it quantum activation
energy. As we show in Sec.~\ref{sec:tunneling}, $R_A$ is smaller
than the tunneling exponent for tunneling between the minima of
$g(P,Q)$. Therefore if the relaxation rate $\eta$ exceeds the
tunneling rate, switching between the stable states occurs via
quantum activation.

Quantum activation energy $R_A$ obtained by solving Eq.
(\ref{eq:BEP}) numerically for $R'$ is plotted in Fig.
\ref{fig:ActE2}. It depends on the control parameter $\mu$ and the
Planck number $\bar{n}$, and it monotonically increases with
increasing $\mu$ and decreasing $\bar{n}$. Close to the
bifurcation point $\mu_B=-1$, i.e., for $\mu-\mu_B\ll 1$, it
displays scaling behavior with $\mu-\mu_B$. From
Eq.~(\ref{eq:Rc0})
\begin{equation}
\label{eq:R_A_bif} R_A=\frac{1}{2}\left(2\bar{n}+1
\right)^{-1}(\mu-\mu_B)^{\zeta},\quad \zeta=2.
\end{equation}
The scaling exponent $\zeta=2$ coincides with the scaling exponent
near the bifurcation point of a classical parametric oscillator
where the oscillator motion is still underdamped in the rotating
frame, i.e., $\mu-\mu_B$ is not to small \cite{Dykman1998}. It can
be seen from the results of
Refs.~\onlinecite{Knobloch1983,Graham1987a} that the exponent
$\zeta=2$ also describes scaling of the activation energy of escape
due to classical fluctuations closer to the pitchfork bifurcation
point where the motion is necessarily overdamped.

\begin{figure}[t]
\includegraphics[width=2.5in]{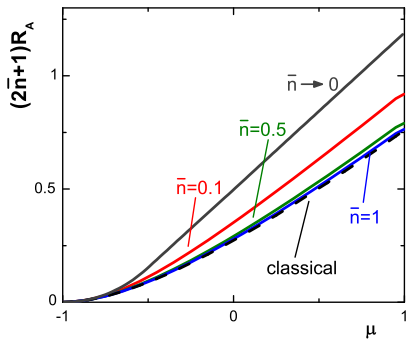}
\caption{(Color online) Quantum activation energy of switching
between the states of parametrically excited vibrations for
different oscillator Planck numbers $\bar n$ as a function of the
scaled frequency detuning $\mu$. The transition rate is $W_{\rm
sw}\propto\exp(-R_A/\lambda)$. With increasing $\bar n$, the value
of $R_A$ multiplied by $2\bar n+1$ quickly approaches the classical
limit $\bar n\gg 1$ shown by the dashed line. In this limit the
ratio $R_A/\lambda$ is $\propto T^{-1}$ and does not contain
$\hbar$. } \label{fig:T_scaled}
\end{figure}

In the classical limit $2\bar{n}+1\gg 1$ we have from
Eq.~(\ref{eq:RC}) $R_A\propto(2\bar{n}+1)\propto T^{-1}$, i.e., the
switching rate $W_{\rm sw}\propto \exp\left[-\left(R_A k_B
T/\lambda\right)/k_B T\right]$, with temperature independent $R_A
k_B T/\lambda$ being the standard activation energy. The quantity
$R_A(2\bar{n}+1)$ in the classical limit as obtained from
Eq.~(\ref{eq:RC}), is shown with the dashed line in
Fig.~\ref{fig:T_scaled}. It is seen that $R_A$ quickly approaches
the classical limit with increasing Planck number $\bar{n}$, so that
even for $\bar{n}=0.1$ the difference between $R_A(2\bar{n}+1)$ and
its classical limit is $\lesssim 15\%$. Effectively it means that,
for small $\bar{n}\ll 1$, the exponent in the probability of quantum
activation can be approximated by the classical exponent for
activated switching in which one should replace
\begin{equation}
\label{eq:T_replacement} k_B T\rightarrow \hbar\omega_0\,.
\end{equation}

\section{Fragility of the detailed balance solution}
\label{sec:Fragility}

It turns out that the expression for the distribution
(\ref{eq:Rho}) and (\ref{eq:R0}) found from the detailed balance
condition for $T=0$ does not generally apply even for
infinitesimally small but nonzero temperature, in the
semiclassical limit. This is a consequence of this solution being
of singular nature, in some sense. A periodically modulated
oscillator should not have detailed balance, because the
underlying time reversibility is broken; the detailed balance
condition (\ref{eq:detailed_balance_condition}) is satisfied just
for one value of $\bar n$ and when other relaxation mechanisms are
disregarded.

 Formally, in a broad range of $\mu$ the correction $\propto
\bar{n}$ to the $T=0$ solution diverges. The divergence can be
seen from Eqs.~(\ref{eq:aE}), (\ref{eq:WMN}), (\ref{eq:R0}). The
transition probabilities $W_{n\,n+m}$ have terms $\propto \bar{n}$
which vary with $m$ as $\exp\left[-2 m\; \im\,\phi_*\right]$ for
$m \gg 1$. At the same time, the $T=0$ solution (\ref{eq:R0})
gives $\xi^{-m}=\exp\left[m\,\omega(g)R'(g)\right]=
\exp\left[2\,m\;\im(\phi_0-2\phi_*)\right]$. Therefore for the
series (\ref{eq:BEP}) with the term $\propto \bar n$ in
$W_{n\,n+m}$ to converge we have to have
\begin{equation}
\label{eq:con}
\im\left(\phi_0-3\phi_*\right)<0\, .
\end{equation}
The condition (\ref{eq:con}) is met at the bottom of the wells of
$g(P,Q)$ and also close to the bifurcation values of the control
parameter, $\mu-\mu_B\ll 1$. We found that, with increasing $\mu$,
the condition (\ref{eq:con}) is broken first for $g$ approaching
the barrier top, $g\rightarrow 0$. In this region $\im\; \phi_0,
\im \;\phi_*\propto \bigl|\ln|g|\bigr|^{-1}$. A somewhat tedious
calculation based on the properties of the elliptic functions
shows that the condition (\ref{eq:con}) is violated when
$\mu>-1/2$, and for $\mu=-1/2$ we have
$\im\,(\phi_0-3\phi_*)\rightarrow 0$ for $g\rightarrow 0$. The
increase in $\mu$ leads to an increase of the range of $g$ where
Eq.~(\ref{eq:con}) does not apply. The detailed balance
distribution is inapplicable for $T\rightarrow 0$ in this range.

To find the distribution for small Planck number $\bar{n}$ in the
range where $\im\left(\phi_0-3\phi_*\right)>0$ we seek the solution
of the balance equation (\ref{eq:BEP}) in the form
\begin{equation}
\label{eq:Rapp}
R'(g)=2\omega^{-1}(g)[\im\, \phi_*(g)-\epsilon]\, ,\,\epsilon\ll
1\, .
\end{equation}
This solution does not  give diverging terms for $\bar{n}>0$. The
terms $W_{n\,n+m}\exp[m\omega(g)R'(g)]$ in Eq.~(\ref{eq:BEP}) are
$\propto \bar n\exp(-2m\epsilon)$ for $m\gg 1$, and their sum over
$m$ is $\propto\bar{n}/\epsilon$. Using the explicit expression for
the transition probabilities we obtain that, to leading order in
$\bar{n}$
\begin{widetext}
\begin{equation}
\label{eq:eps} \epsilon=\bar{n}\left[\sum_m
\exp\left[m\;\im(\phi_0-3\phi_*)\right]\sinh\left(m\;\im\,\phi_*\right)/
\left|\cos\left(m\;\im \,\phi_0/2\right)\right|^2\right]^{-1}.
\end{equation}
\end{widetext}
Here again, the sum goes over positive and negative $m$. In
obtaining Eq.~(\ref{eq:eps})  we used the general expression
(\ref{eq:am}) for the coefficients $a_m(g)$ in $W_{n+m\,n}$. In
fact, the asymptotic expressions (\ref{eq:aE}) for $a_m(g)$ allow
one to calculate the sum over $m$ in Eq.~(\ref{eq:eps}) explicitly
and the result provides a very good approximation for $\epsilon$.

Equations (\ref{eq:Rapp}) and (\ref{eq:eps}) give the effective
inverse temperature $R'/\lambda$ and therefore the distribution
$\rho_n$ as a whole for $\bar{n}\ll 1$. It is clear that the
$\bar{n}\rightarrow 0$ limit is completely different from the
detailed balance solution (\ref{eq:R0}) for $\bar{n}=0$, that is,
the transition to the $T=\bar{n}=0$ regime is nonanalytic. The
difference between the $\bar{n}=0$ and the $\bar{n}\rightarrow 0$
solution is seen in Figs.~\ref{fig:Fr} and \ref{fig:ActE2}.

It follows from Eq.~(\ref{eq:eps}) that the perturbation theory
diverges for the value of $g$ where
$\im\left(\phi_0-3\phi_*\right)=0$. At such $g$ the
$\bar{n}\rightarrow 0$ solution (\ref{eq:Rapp}) coincides with the
$\bar{n}=0$ solution (\ref{eq:R0}). For smaller $g$ the condition
(\ref{eq:con}) is satisfied and the $T=0$ solution for $R'(g)$
applies; the corrections to this solution are $\propto \bar{n}\ll
1$. The derivative of the effective inverse temperature
$R'(g)/\lambda$ over $g$ is discontinuous at the crossover between
$\bar{n}\rightarrow 0$ and $\bar{n}=0$ solutions, as seen in
Fig.~\ref{fig:Fr}. Figure \ref{fig:Fr} illustrates also the
smearing of the singularity of $R'(g)$ due to terms $\propto \bar
n$ in $W_{nn'}$, which is described by the numerical solution of
the balance equation (\ref{eq:BEP}). Away from the crossover the
analytical solutions provide a good approximation to numerical
results.

\subsection{Breaking of detailed balance solution by dephasing}

Dephasing plays an important role in the dynamics of quantum
systems. It comes from fluctuations of the transition frequency due
to external noise or to coupling to a thermal reservoir. A simple
mechanism is quasielastic scattering of excitations of the reservoir
off the quantum system. Since the scattering amplitude depends on
the state of the system, the scattering leads to diffusion of the
phase difference of different states.

For an oscillator, dephasing has been carefully studied, both
microscopically and phenomenologically, see
Refs.~\onlinecite{Ivanov1966a,Barker1975,DK_review84} and papers
cited therein. It leads to an extra term $-\hat{\eta}^{\rm ph}\rho$
in the quantum kinetic equation (\ref{eq:Master}), with
\begin{equation}
\label{eq:dephasing} \hat{\eta}^{\rm ph}\rho=\eta^{\rm
ph}\left[\had\ha,\left[\had\ha,\rho\right]\right],
\end{equation}
where $\eta^{\rm ph}$ is the dimensionless dephasing rate.

If both $\eta_{\rm ph}$ and $\eta$ are small compared to
$\omega(g)$, populations of the steady states $\rho_n$ are described
by the balance equation (\ref{eq:BE}) in which one should replace
$\eta W_{n'n}\rightarrow\eta W_{n'n}+\eta^{\rm ph} W_{n'n}^{\rm
(ph)}$, with
\begin{equation}
\label{eq:W_dephasing} W_{n'n}^{\rm (ph)}=\left|\lz
n|\had\ha|n'\rz\right|^2.
\end{equation}
It follows from Eq.~(\ref{eq:aE}) that, for large $|n'-n|$, we have
$W_{n'n}^{\rm (ph)}\propto\exp\left[-2|n'-n|\im \,\phi_*\right]$,
that is, the transition probability exponentially decays with
increasing $|n'-n|$, and the exponent is determined by
$\im\,\phi_*$.

Even slow dephasing is sufficient for making the detailed balance
condition inapplicable. Mathematically, the effect of slow dephasing
is similar to the effect of nonzero temperature. If the condition
(\ref{eq:con}) is violated, the sum $\sum W_{n+m\,n}^{\rm
(ph)}\exp\left(-m\omega(g_n)R'(g_n)\right)$ with $R'(g)$ given by
the detailed balance solution (\ref{eq:R0}) diverges. The correct
distribution for the appropriate $g$ and $\mu$ is given by
Eq.~(\ref{eq:Rapp}). The parameter $\epsilon\equiv \epsilon(g)$ is
given by Eq. (\ref{eq:eps}) in which $\bar{n}$ is replaced,
\begin{eqnarray}\label{eq:ePh}
&&\bar{n}\rightarrow\bar{n}+C^{\rm ph}\eta^{\rm ph}/\eta,\qquad
C^{\rm ph}=\omega^{2}(g)/2\lambda
\nonumber\\
&&\times\left|\sum_{m=0}^{\infty}\,\exp\left(2im
\phi_*\right)/\left[\exp\left(im\phi_0\right)+1\right]\right|^2.
\end{eqnarray}
It is seen from Eq.~(\ref{eq:ePh}) and Figs. \ref{fig:Fr} and
\ref{fig:ActE2} that, for low temperatures, even weak dephasing,
$\eta^{\rm ph}/\eta\ll 1$, leads to a very strong change of the
probability distribution.

The fragility of the detailed balance solution discussed in this
section is a semiclassical effect. It occurs if the number of states
in each well $N\propto \lambda^{-1}$ is large. Formally we need
$\epsilon\exp[2cN\,{\rm Im}\,(\phi_0-3\phi_*)]\gtrsim 1$, with
$c=c(g)\sim 1$. In other words, the detailed balance solution is
fragile provided $\lambda$ is sufficiently small. A full numerical
solution of the balance equation confirms that, when $\lambda$ is no
longer small, the distribution for small $\bar n, \eta^{\rm
ph}/\eta$ is close to that described by the $\bar n=0, \eta^{\rm
ph}/\eta =0$.

\section{Tunneling}
\label{sec:tunneling}

 The oscillator localized initially in one of the wells of the
effective Hamiltonian $g(P,Q)$ can switch to another well via
tunneling. For small $\lambda$ tunneling can be described in the WKB
approximation. We will first find the tunneling exponent assuming
that the oscillator is in the lowest intrawell state and show that
it exceeds the quantum activation exponent $R_A/\lambda$. We will
then use standard arguments to show that this is true independent of
the initially occupied intrawell state, for all temperatures.

If we now go back to the original problem of switching between
stable states of a periodically modulated oscillator, we see that
switching via tunneling can be observed only where the relaxation
rate is smaller than the tunneling rate, that is the prefactor in
the rate of quantum activation is very small. Such experiment
requires preparing the system in one of the wells. Tunneling between
period two states of a strongly modulated strongly nonlinear system
has been demonstrated in atomic optics
\cite{Hensinger2001,Steck2001,Hensinger2003}. For a weakly nonlinear
oscillator, tunneling splitting between the lowest states of the
Hamiltonian $\hat g$ for $\mu=0$ was found in
Ref.~\cite{Wielinga1993}.

The effective Hamiltonian $g(P,Q)$ is quartic in the momentum $P$.
Therefore the semiclassical momentum $P(Q,g)$ as given by the
equation $g(P,Q)=g$ has 4 rather than 2 branches. This leads to new
features of tunneling compared to the standard picture for
one-dimensional systems with a time-independent Hamiltonian
quadratic in $P$.

For concreteness, we will consider tunneling from the left well,
which is located at $Q=Q_{l0}=-(1+\mu)^{1/2}, P=0$. The rate of
tunneling from the bottom of the well of $g(P,Q)$ in
Fig.~\ref{fig:g} is determined by the WKB wave function with
$g=g_{\min}=-(1+\mu)^2/4$. This wave function is particularly simple
for $\mu < 0$. In the region $Q_{l0}< Q$ and not too close to
$Q_{l0}$ it has the form
\begin{eqnarray}
\label{eq:tail}&& |\psi_l\rangle =C(\partial_P
g)^{-1/2}\exp[iS_0(Q)/\lambda],\nonumber\\
&&S_0(Q)=\int\nolimits_{Q_{l0}}^Q\,P_-(Q')\,dQ',
\end{eqnarray}
where
\begin{equation}
\label{eq:P_pm} P_{\pm}(Q)=i\left[1 \pm
\left(Q^2-\mu\right)^{1/2}\right].
\end{equation}

For $\mu<0$ the wave function $|\psi_l\rangle$ monotonically decays
with increasing $Q$. The exponent for interwell tunneling is $S_{\rm
tun}/\lambda$, with $S_{\rm tun}= {\rm Im}\;S_0(-Q_{l0})$ (we use
that $-Q_{l0}$ is the position of the right well),
\begin{equation}
\label{eq:S_tun} S_{\rm tun}=
(1+\mu)^{1/2}+\mu\log\frac{1+(1+\mu)^{1/2}}{|\mu|^{1/2}}.
\end{equation}

A more interesting situation arises in the case $\mu > 0$. Here,
inside the classically forbidden region $-\mu^{1/2} <Q< \mu^{1/2}$
decay of the wave function is accompanied by oscillations. We
leave the analysis of this behavior for a separate paper. Here we
only note that the tunneling exponent is still given by
Eq.~(\ref{eq:S_tun}).

\begin{figure}[h]
\begin{center}
\includegraphics[width=3.0in]{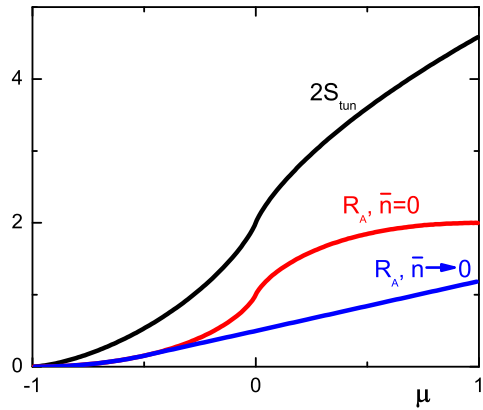}
\caption{(Color online) The scaled exponent $2S_{\rm tun}$ in the
tunneling probability as a function of the parameter
$\mu=2\omega_F\delta\omega/F$ (solid line). Also shown for
comparison is the quantum activation energy $R_A$ for $\bar n=0$ and
$\bar n\to 0$.}\label{fig:tunneling}
\end{center}
\end{figure}

If the level splitting due to tunneling is small
compared to their broadening due to relaxation, the tunneling
probability is quadratic in the tunneling amplitude,
\begin{equation}
\label{eq:tunnel_switching} W_{\rm tun}\propto \exp(-2S_{\rm
tun}/\lambda).
\end{equation}
The action $2S_{\rm tun}$ as a function of $\mu$ is plotted in
Fig.~\ref{fig:tunneling}. It is seen from this figure that the
tunneling exponent exceeds the quantum activation exponent
$R_A/\lambda$ for all values of the control parameter $\mu$. This
indicates that, as mentioned above, it is exponentially more
probable to switch between the classically stable states of the
oscillator via activation than via tunneling from the ground
intrawell state, for not too small relaxation rate.

In the same limit where the relaxation rate exceeds the tunneling
rate, we can consider the effect of tunneling from excited
intrawell states of the Hamiltonian $\hat g$. The analysis is
similar to that for systems in thermal equilibrium
\cite{Affleck1981,Grabert1984,Larkin1985}. Over the relaxation
time there is formed a quasiequilibrium distribution over the
states inside the initially occupied well of $g(P,Q)$. As before,
we assume that relaxational broadening of the levels $g_n$ is
small compared to the level spacing. The probability of tunneling
from a given state $n$ is determined by its occupation $\rho_n$.
The overall switching probability is given by a sum of tunneling
probabilities from individual intrawell states
\begin{equation}
\label{eq:multilevel_W} W_{\rm sw}=\sum_nC_ne^{-2S_n/\lambda}\rho_n,
\end{equation}
where $C_n$ is a prefactor that smoothly depends on $n$, and
\[S_n={\rm Im}\,\int P(Q,g_n)dQ\]
is the imaginary part of the action of a classical particle with
Hamiltonian $g(P,Q)$ and energy $g_n$, which moves in complex time
from the turning point $P=0$ in one well of $g(P,Q)$ to the
turning point in the other well. It follows from the analysis in
Appendix~A and Eq.~(\ref{eq:Pe}), that the duration of such motion
is $(\tau_p^{(1)} + \tau_p^{(2)})/2$. Therefore $\partial
S_n/\partial g_n = -{\rm Im}\,\tau_p^{(2)}/2$.

Taking into account that $\rho_n=\exp(-R_n/\lambda)$, we see that
the derivative of the overall exponent in
Eq.~(\ref{eq:multilevel_W}) over $g_n$ is ${\rm
Im}\,\tau_p^{(2)}-R'(g_n)$. It was shown in
Secs.~\ref{sec:Distribution} and \ref{sec:Limiting_cases} that
this derivative is always positive, cf. Fig.~\ref{fig:Randt}.
Therefore the exponent monotonically increases (decreases in
absolute value) with increasing $g$. This shows that, with
overwhelming probability, switching occurs via overbarrier
transitions, i.e., the switching mechanism is quantum activation.
We emphasize that this result is independent of the bath
temperature.

\section{Conclusions}
\label{sec:Conclusions}

In this paper we studied switching between the states of period two
vibrations of a parametrically modulated nonlinear oscillator and
the distribution over period-two quasienergy levels. We considered a
semiclassical case where the wells of the scaled oscillator
Hamiltonian in the rotating frame $g(P,Q)$, Fig.~\ref{fig:g},
contain many levels. The distance between the levels is small
compared to $\hbar\omega_F$ but much larger than the tunneling
splitting. We assumed that the oscillator is underdamped, so that
the interlevel distance exceeds their width. At the same time, of
primary interest was the case where this width exceeds the tunneling
splitting.

We considered relaxation due to coupling to a thermal bath, which is
linear in the oscillator coordinate, as well as dephasing from
random noise that modulates the oscillator frequency. The problem of
the distribution over intrawell states was reduced to a balance
equation. The coefficients in this equation were obtained explicitly
in the WKB approximation, using the analytical properties of the
solution of the classical equations of motion.  The balance equation
was then solved in the eikonal approximation. The eikonal solution
was confirmed by a full numerical solution of the balance equation,
which did not use the WKB approximation for transition matrix
elements.

We found that the distribution over period-two quasienergy has a
form of the Boltzmann distribution with effective temperature that
depends on the quasienergy. This temperature remains nonzero even
where the temperature of the thermal bath $T\to 0$. It is determined
by diffusion over quasienergy, which accompanies relaxation and has
quantum origin: it is due to the Floquet wave functions being
combinations of the Fock wave functions of the oscillator.

Unexpectedly, we found that the quasienergy distribution for $T=0$,
where the system has detailed balance, is fragile. It differs
significantly from the solution for $T\to 0$, for a large number of
intrawell states. The $T=0$ solution is also destroyed by even small
dephasing.

The probability of switching between period two states $W_{\rm
sw}$ is determined by the occupation of the states near the
barrier top of the effective Hamiltonian $g(P,Q)$. We calculated
the effective quantum activation energy $R_A$ which gives $W_{\rm
sw}\propto \exp(-R_A/\lambda)$. Both $R_A/\lambda$ and the
exponent of the tunneling probability are proportional to the
reciprocal scaled Planck constant $\lambda^{-1}$. However, for all
parameter values and all bath temperatures, $R_A/\lambda$ is
smaller than the tunneling exponent. Therefore in the case where
intrawell relaxation is faster than interwell tunneling, switching
occurs via quantum activation.

In the limit where fluctuations of the oscillator are classical,
$kT\gg \hbar\omega_F$, we have $R_A\propto (kT)^{-1}$, and $W_{\rm
sw}$ is described by the standard activation law. Down to small
Planck numbers $\bar n\gtrsim 0.1$ the quantum activation energy
$R_A$ is reasonably well described by the classical expression even
for small $kT/\hbar\omega_F$ provided $kT$ is replaced by
$\hbar\omega_F(2\bar n+1)/4$, with $\bar n$ being the Planck number
of the oscillator. The inapplicability of this description for small
$\bar n$ indicates, however, that classical and quantum fluctuations
do not simply add up. The replacement $kT\to \hbar\omega_F(2\bar
n+1)/4$ becomes exact for all $\bar n$ close to the
bifurcational value of the control parameter $\mu=\mu_B$ where the
period-two states first emerge. In this range $R_A$ scales with the
distance to the bifurcation point as $R_A\propto (\mu-\mu_B)^2$.

The results on switching rate are accessible to direct
experimental studies in currently studied nano- and microsystems,
in particular in systems based on Josephson junctions, including
those used for highly sensitive quantum measurements.

We gratefully acknowledge a discussion with M. Devoret. This work
was partly supported by the NSF through grant No. ITR-0085922.

\appendix
\section{Classical motion of the parametric oscillator}\label{ap:CM}

The Hamiltonian $g(P,Q)$ (\ref{eq:g}) is quartic in $P$ and $Q$.
This makes it possible to solve classical equations of motion
(\ref{eq:dQPdt}). Trajectories with given $g$ lie on the
cross-section $g(P,Q)=g$ of the surface $g(P,Q)$ in Fig.
\ref{fig:g}. We will be interested only in the intrawell
trajectories, in which case $g\leq 0$. The trajectories in different
wells are inversely symmetrical and can be obtained by the
transformation  $Q\rightarrow -Q$, $P\rightarrow -P$. Their time
dependence can be expressed in terms of the Jacobi elliptic
functions \cite{Abramowitz1972}. For trajectories in the right well
in Fig.~\ref{fig:g}, where $Q>0$, we have
\begin{eqnarray}\label{eq:QPt}
Q(\tau)=\frac{2^{3/2}|g|^{1/2}\dn\,\tau'}{\kp+\km \cn\,\tau'}
,\\
P(\tau)=\frac{\kp\km|g|^{1/4}\sn\,\tau'}{\kp+\km\cn\,\tau'}
.\nonumber
\end{eqnarray}
Here
\begin{equation}
\label{eq:kappa}
\kappa_{\pm}=\left(1+\mu\pm
2|g|^{1/2}\right)^{1/2},\quad \tau'=2^{3/2}|g|^{1/4}\tau .
\end{equation}
The parameter of the elliptic functions $m_J=m_J(g)$ is given by
Eq.~(\ref{eq:m}). For $\mu<0$ the function $m_J(g)$ monotonically
decreases with increasing $g$ from $m_J=0$ for
$g=g_{\min}=-(\mu+1)^2/4$ to $m_J\to -\infty$ for $g\to 0$. For
$\mu>0$ the function $m_J(g)$ becomes nonmonotonic. It first
increases from $m_J=0$ with increasing $g$, but than decreases, goes
trough $m_J=0$ for $g=-(1-\mu)^2/4$, and goes to $-\infty$ for
$g\rightarrow 0$. The Jacobi functions $\sn\left(\tau'|m_J\right)$,
$\cn(\tau'|m_J)$ and $\dn(\tau'|m_J)$ for $m_J<0$ are equal to
$(1-m_J)^{-1/2}\sd\left(\tilde{\tau}'|\tilde{m}_J\right)$,
$\cd\left(\tilde{\tau}'|\tilde{m}_J\right)$, and
$\nd\left(\tilde{\tau}'|\tilde{m}_J\right)$ with
$\tilde{m}_J=-m_J/(1-m_J)$ and $\tilde{\tau}'=(1-m_J)^{1/2}\tau'$
\cite{Abramowitz1972}.

The double periodicity of the functions $Q(\tau)$, $P(\tau)$
discussed in Sec. \ref{sec:gspec} is a consequence of the double
periodicity of elliptic functions. The expressions for the periods
$\tau_p^{(1,2)}$, Eq.~(\ref{eq:Period}), follow from
Eq.~(\ref{eq:QPt}). The trajectories in the left well of $g(P,Q)$ in
Fig. \ref{fig:g}, where $Q<0$, can be written in the form
\begin{eqnarray}\label{eq:Pe}
&&
Q_l(\tau)=Q\left(\tau+\left(\tau_p^{(1)}+\tau_p^{(2)}\right)/2\right)\,
,\\
&&
P_l(\tau)=P\left(\tau+\left(\tau_p^{(1)}+\tau_p^{(2)}\right)/2\right).\nonumber
\end{eqnarray}
This expression shows how to make a transition from one well to
another  by moving in complex time, which simplifies the analysis of
oscillator tunneling.

We note that the function $\cn\,\tau'$ in the expressions for
$Q,P$ (\ref{eq:QPt})  has periods $\tau_p^{(1)}$,
$\left(\tau_p^{(1)}+\tau_p^{(2)}\right)/2$ \cite{Abramowitz1972}
as a function of $\tau$. It's period parallelogram is shown in
Fig.~\ref{fig:CL}. In this parallelogram $\cn\,\tau'$ takes on any
value twice. Therefore both $Q(\tau)$ and $P(\tau)$ have two poles
located at $\tau=\tau_*, \tau_{**}$. The values of $\tau_*,
\tau_{**}$ are given by the equation
\begin{eqnarray}
\label{eq:tau*}
&&\cn(2^{3/2}|g|^{1/4}\tau)=-\kp/\km\, ,\\
&&\tau_{**}=\frac{1}{2}\left(3\tau_p^{(1)}+\tau_p^{(2)}\right)-\tau_*\,
.\nonumber
\end{eqnarray}
The positions of the poles are shown in Fig.~\ref{fig:CL} for
$m_J>0$ and $m_J<0$, respectively, with
$\phi_*=2\pi\tau_*/\tau_p^{(1)},
\phi_{**}=2\pi\tau_{**}/\tau_p^{(2)}$. For concreteness we choose
\begin{equation}
\label{eq:tau_choice} \im\,\tau_*<\im\,\tau_p^{(2)}/4.
\end{equation}
Using the relations between the Jacobi elliptic functions
\cite{Abramowitz1972} we find that, near the pole at $\tau=\tau_*$,
\begin{equation}
\label{eq:pole_a} P-iQ\propto -(\tau-\tau_*)^{-1},
\end{equation}
whereas $P-iQ$ is not singular at $\tau=\tau_{**}$.
Eq.~(\ref{eq:pole_a}) was used to obtain the explicit form of the
matrix element of the operator $\ha=(2\lambda)^{-1/2}(P-iQ)$ in
Sec.~II~C.

\section{The classical Limit}\label{ap:C}

In this Appendix we calculate the effective inverse temperature
$R'(g)$ in the limit of large oscillator Planck number $\bar{n}\gg
1$. The explicit form of the coefficients in Eq.~(\ref{eq:CW}) for
$R'$ follows from the general expression (\ref{eq:WMN}) for the
transition rates $W_{nn'}$,
\begin{eqnarray}\label{eq:Sums}
\sum_m m\,W_{n+m\,n}=\sum_{m=-\infty}^{\infty} m|a_{-m}(g_n)|^2\, ,\\
\sum_m m^2\,W_{n+m\,n}=(2\bar{n}+1)\sum_{m=-\infty}^{\infty}
m^2|a_{-m}(g_n)|^2.\nonumber
\end{eqnarray}
The semiclassical matrix elements $a_{-m}$ are given by Eq.
(\ref{eq:a}). They are Fourier components of the function
$a(\phi;g)$ on the classical trajectory with given $g=g_n$.

Using the completeness condition
$\sum\nolimits_{m=-\infty}^{\infty}e^{im(\phi_2-\phi_1)}=
2\pi\delta(\phi_2-\phi_1)$ we can rewrite
\begin{eqnarray}
\label{eq:sums_m}
&&\sum_{m=-\infty}^{\infty}m|a_{-m}(g)|^2=\frac{1}{2i\pi}\int_0^{2\pi}d\phi
\, a(\phi;g)\,\partial_{\phi}a^{*}(\phi;g),\nonumber\\
&&\sum_{m=-\infty}^{\infty}m^2|a_{-m}(g)|^2=(2\pi)^{-1}\nonumber\\
&&\qquad\times\int_0^{2\pi}d\phi\,\partial_{\phi}
a(\phi;g)\,\partial_{\phi}a^{*}(\phi;g),
\end{eqnarray}
with $a(\phi;g)=(2\lambda)^{-1/2}\bigl(P(\tau;g)-iQ(\tau;g)\bigr)$
and $\phi = \omega(g)\tau$. The integrals over $\phi$ can be written
as contour integrals over $dP$, $dQ$ along the trajectories with
given $g$. The contour integrals can be further simplified using the
Stokes theorem. This gives
\begin{eqnarray}
\label{eq:sums_to_MN}
\sum_{m=-\infty}^{\infty}m|a_{-m}(g)|^2 & \equiv & \frac{1}{2\lambda\pi}M(g)\, ,\\
\sum_{m=-\infty}^{\infty}m^2|a_{-m}(g)|^2 & \equiv &
\frac{\omega^{-1}(g)}{2\lambda\pi} N(g)\, ,\nonumber
\end{eqnarray}
where the functions $M(g)$ and $N(g)$ are given by Eq.
(\ref{eq:RC}).



\end{document}